\documentclass[twocolumn,preprintnumbers,amsmath,superscriptaddress,amssymb,aps]{revtex4-1}

\usepackage{graphicx}
\usepackage{dcolumn}
\usepackage{bm}
\usepackage{epsfig}
\usepackage{amsmath}
\usepackage{amssymb}
\usepackage{graphicx}
\usepackage{indentfirst}
\usepackage{booktabs}
\usepackage{multirow}
\usepackage{colortbl}

\def\degree{${}^{\circ}$}
\selectfont

\begin{document}

\title{A First Principle Study on Magneto-Optical Effects in Ferromagnetic Semiconductors Y$_3$Fe$_5$O$_{12}$ and Bi$_3$Fe$_5$O$_{12}$}
\author{Wei-Kuo Li}
\address{Department of Physics and Center for Theoretical Physics, National Taiwan University, Taipei 10617, Taiwan}
\author{Guang-Yu Guo}
\email{gyguo@phys.ntu.edu.tw}
\address{Department of Physics and Center for Theoretical Physics, National Taiwan University, Taipei 10617, Taiwan}
\address{Physics Division, National Center for Theoretical Sciences, Hsinchu 30013, Taiwan}

\date{\today}

\begin{abstract}
The magneto-optical (MO) effects not only are a powerful probe of 
magnetism and electronic structure of magnetic solids 
but also have valuable applications in high-density data-storage technology. 
Yttrium iron garnet (Y$_3$Fe$_5$O$_{12}$) (YIG) and bismuth iron garnet 
(Bi$_3$Fe$_5$O$_{12}$) (BIG) are two widely used magnetic semiconductors 
with significant magneto-optical effects.
In particular, YIG has been routinely used as a spin current injector. 
In this paper, we present a thorough theoretical investigation on 
magnetism, electronic, optical and MO properties of YIG and BIG, 
based on the density functional theory with the generalized gradient approximation plus onsite Coulomb repulsion. 
We find that YIG exhibits significant MO Kerr and Faraday effects in UV frequency range that are
comparable to ferromagnetic iron. Strikingly, BIG shows gigantic MO effects in visible
frequency region that are several times larger than YIG.
We find that these distinctly different MO properties of
YIG and BIG result from the fact that the magnitude of
the calculated MO conductivity ($\sigma_{xy}$) of BIG is one order of magnitude larger than that of YIG.
Interestingly, the calculated band structures reveal that both valence and 
conduction bands across the semiconducting band gap in BIG are purely spin-down states, 
i.e., BIG is a single spin semiconductor.
They also show that in YIG, Y $sd$ orbitals mix mainly with the high lying 
conduction bands, leaving Fe $d$ orbital dominated lower conduction
bands almost unaffected by the SOC on the Y atom. 
In contrast, Bi $p$ orbitals in BIG hybridize significantly with Fe $d$ orbitals in
the lower conduction bands, leading to large SOC-induced band splitting in the bands. 
Consequently, the MO transitions between the upper valence bands and lower conduction bands 
are greatly enhanced when Y is replaced by heavier Bi. 
This finding suggests a guideline in search for materials with desired MO effects.
Our calculated Kerr and Faraday rotation angles of YIG agree well with the available experimental values.
Our calculated Faraday rotation angles for BIG are in nearly perfect
agreement with the measured ones. Thus, we hope that
our predicted giant MO Kerr effect in BIG will stimulate further MOKE experiments on high quality BIG crystals.
Our interesting findings show that the iron garnets
not only offer an useful platform for exploring the interplay of microwave, spin current, magnetism, 
and optics degrees of freedom, but also have
promising applications in high density MO data-storage and low-power consumption spintronic nanodevices.
\end{abstract}

\maketitle

\section{Introduction}
Yttrium iron garnet (Y$_3$Fe$_5$O$_{12}$, YIG) is a 
ferrimagnetic semiconductor with excellent magnetic properties 
such as high curie temperature $T_c$~\cite{cherepanov1993}, low Gilbert 
damping $\alpha \sim6.7\times10^{-5}$~\cite{kajiwara2010,mizukami2002,chikazumi1997} 
and long spin wave propragating length~\cite{schneider2008}. 
Various applications such as spin pumping require a non-metallic magnet.  
YIG is thus routinely used for spin pumping purposes~\cite{kajiwara2010}. 
It is also widely used as a 
magnetic insulating substrate for purposes such as introducing 
magnetic proximity effect while avoiding electrical short-cut.~\cite{sun2013} YIG has high Curie temperature, 
which is good for applications across a wide temperature range. 
The low Gilbert damping of YIG also makes it a good microwave material. 
YIG thus becomes a famous material in the field of spintronics, 
where coupling between magnetism, microwave and spin current becomes possible.  

Magneto-optical (MO) effects are important examples of light-matter interactions in magnetic phases.~\cite{oppeneer2001,antonov2004}   
When a linearly polarized light beam is shined onto a magnetic material, the reflected and transmitted light 
becomes elliptically polarized. The principal axis is rotated with respect to the polarization direction 
of incident light beam. The former and latter effects are termed MO Kerr (MOKE) and MO Faraday (MOFE) effects, 
respectively. MOKE  allowes us to detect the magnetization locally with a high spatial and temporal resolution 
in a non-invasive fashion. Furthermore, magnetic materials with large MOKE would find valuable MO storage 
and sensor applications~\cite{mansuripur1995,castera1996}. Thus it has been widely used to probe the electronic 
and magnetic properties of solids, surface, thin films and 2D magnets~\cite{antonov2004}. On the other hand, 
MOFE can be used as a time-reversal symmetry-breaking element in optics~\cite{haldane2008}, 
and its applications such as optical isolators are consequenses of  time-reversal symmetry-breaking~\cite{aplet1964}. 
Magnetic materials with large Kerr or Faraday rotation angles have technological applications. 

YIG is also known to be MO active ~\cite{dillon1958}. Various experiments have been carried out to study 
the MOKE and MOFE of iron garnets in the visible and near-UV regime ~\cite{wittekoek1975, kahn1969}. 
Substituting yttrium with bismuth results in bismuth iron garnet (Bi$_3$Fe$_5$O$_{12}$) (BIG). BIG has approximately 
7 times larger Faraday rotation angles than that of YIG. The effect of doping bismuth into YIG on the  
MOFE spectrum was studied~\cite{chern1999, jesenska2016}. The large radius of bismuth atoms seems to make bulk BIG 
unstable. Thus high quality BIG film is difficult to synthesize~\cite{vertruyen2008}. 
Though numerous experimental studies have been done on these systems, first-principle calculations are scarce. 
This is probably due to the complexity of the structures of BIG and YIG. As shown in Fig.  1(a), 
they have a total of 80 atoms in the primitive cell. Although the electronic structures of YIG and BIG 
have been theoretically studied~\cite{xu2000,oikawa2005},
no first principle calculation on the MOKE or MOFE spectra of YIG and BIG have been reported. 
Therefore, here we carry out a systematic first-principle density functional study 
on the optical and MO properties of YIG and BIG. 
The rest of this paper is organized as follows. 
A brief description of the crystal structures of YIG and BIG as well as the theoretical methods used
is given in Sec. II. In Sec. III, the calculated magnetic moments, electronic structure, 
optical conductivities, MO Kerr and Faraday effects
are presented. Finally, the conclusions drawn from this work are given in section IV.  

\section{CRYSTAL STRUCTURE AND COMPUTATIONAL METHODS }
YIG and BIG crystalize in the cubic structure with space group 
$Ia3d$~\cite{bertaut1956, toraya1995}, as illustrated in Fig. 1(a). 
In each unit cell, there are 48 oxygen atoms at the Wyckoff 96h positions, 
8 octahedrally coordinated iron atoms (Fe$^O$) 
at the 16a positions, and 12 tetrahedrally coordinated iron atoms (Fe$^T$) 
at the 24d positions in the primitive cell.
In other words, there are two Fe$^O$ ions and three Fe$^T$ ions per formula unit (f.u.).  
The experimental lattice constant $a=12.376$ \AA, and the experimental Wyckoff parameters for oxygen atoms 
are $(x, y, z) = (0.9726, 0.0572, 0.1492)$.~\cite{bertaut1956} 
The experimental lattice constant for BIG $a=12.6469$ \AA.~\cite{toraya1995} 
Accurate oxygen position measurement 
for BIG is still on demand and under debate~\cite{vertruyen2008}. 
Therefore we use the experimental lattice constant for BIG 
with the atomic positions determined theoretically (see Table I), as described next.
We use the experimental lattice constant and atomic positions for all YIG calculations,

\begin{table}[]
\caption{Structural parameters of Y$_3$Fe$_5$O$_{12}$ and Bi$_3$Fe$_5$O$_{12}$. 
For YIG, experimental lattice constant $a=12.376$ \AA$ $ and oxygen positions~\cite{bertaut1956} are used. 
For BIG, experimental lattice constant $a=12.6469$ \AA~\cite{toraya1995} is used while the oxygen positions 
are determined theoretically.}
\begin{tabular}{lllll}
\hline\hline
Y$_3$Fe$_5$O$_{12}$   & Wyckoff position  & $x$     &$y$       &$z$     \\ \hline
Fe$^O$               & 16a              & 0.0000 & 0.0000   & 0.0000 \\
Fe$^T$               & 24d              & 0.3750 & 0.0000   & 0.2500 \\
Y                    & 24c              & 0.1250 & 0.0000   & 0.2500 \\
O                    & 96h              & 0.9726\footnotemark[1] & 0.0572\footnotemark[1]  & 0.1492\footnotemark[1]  \\ \hline
Bi$_3$Fe$_5$O$_{12}$  & Wyckoff position  & $x$     &$y$       &$z$     \\ \hline
Fe$^O$               & 16a              & 0.0000 & 0.0000   & 0.0000 \\
Fe$^T$               & 24d              & 0.3750 & 0.0000   & 0.2500 \\
Bi                   & 24c              & 0.1250 & 0.0000   & 0.2500 \\
O                    & 96h              & 0.0540 & 0.0300 & 0.1485 \\ \hline \hline
\end{tabular}
\footnotemark[1]{Ref.~\onlinecite{bertaut1956}.}
\end{table}

\begin{figure}
\centerline{\psfig{file=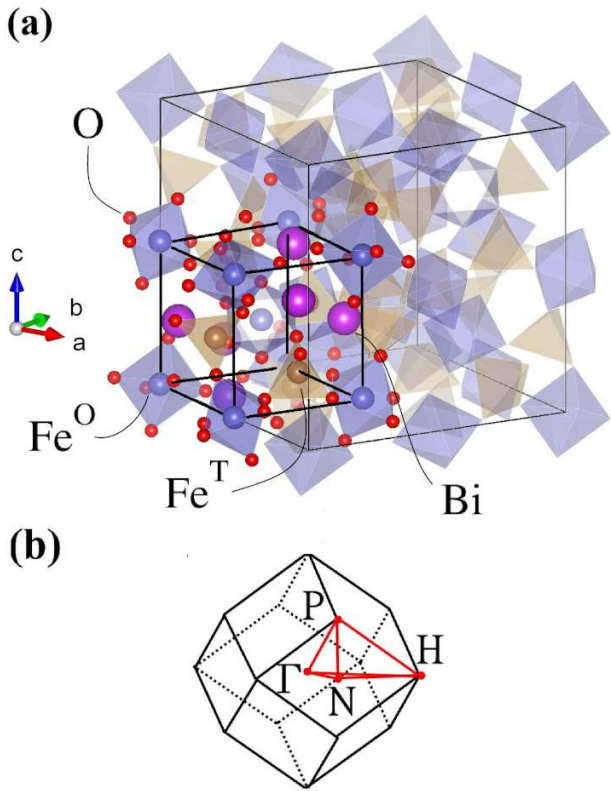,width=0.96\linewidth}}
\caption{(a) 1/8 of BIG conventional unit cell. Oxygen atoms are shown as red balls; 
bismuth atoms are shown as purple balls; Fe$^T$ atoms are shown as yellow balls; Fe$^O$ atoms are shown as blue balls. 
(b) Brillouin zone of both YIG and BIG. The red lines denote the high symmetry lines where the calculated
energy bands will be plotted.}
\label{fig.1}
\end{figure}

Our first principle calculations are based on the density functional theory 
with the generalized gradient approximation (GGA) 
of the Perdew-Burke-Ernzerhof formula~\cite{perdew1996} 
to the electron exchange-correlation potential. 
Furthermore, we use the  GGA + $U$ method to have a better 
description for on-site interaction for Fe $d$ electrons.~\cite{dudarev1998}  
Here we set $U = 4.0$ eV, which was found to be rather 
appropriate for iron oxides~\cite{jeng2004}.
Indeed, as we will show below, the optical and MO spectra calculated 
using this $U$ value agree rather well with the available experimental spectra.
All the calculations are carried out by using the accurate 
projector-augmented wave~\cite{kresse1999} method, 
as implemented in Vienna \textit{ab initio} Simulation Package 
(VASP).~\cite{kresse1996a,kresse1996b} 
A large energy cutoff of 450 eV for the plane-wave basis is used. 
A \(6 \times 6 \times 6\) $k$-point mesh is used 
for both systems in the self-consistent charge density calculations. 
The density of states (DOS) calculation 
is performed with a denser $k$-point mesh of \(10 \times 10 \times  10\).

We first calculate the optical conductivity tensor which determine the MOKE and MOFE.
We let the magnetization of our systems be along (001) ($z$) direction. 
In this case, our systems have the four-fold rotational symmetry along the $z$ axis 
and thus the optical conductivity tensor can be written in the following form~\cite{feng2015}:
\begin{equation}
\sigma=
\begin{pmatrix}
 \sigma_{xx} & \sigma_{xy}   & 0 \\
 -\sigma_{xy}& \sigma_{xx} & 0 \\
 0 & 0  & \sigma_{zz}
\end{pmatrix}.
\end{equation}
The optical conductivity tensor can be formulated within the linear response theory. 
Here the real part of the diagonal elements and 
imaginary part of the off-diagonal elements
are given by~\cite{wang1974,oppeneer1992,feng2015}:
\begin{equation}
\sigma_{aa}^{1} (\omega) = \frac{\pi e^2}{\hbar\omega m^2}
\sum_{i,j}\int_{BZ}\frac{d{\bf k}}{(2\pi)^3}|p_{ij}^{a}|^{2}
\delta(\epsilon_{{\bf k}j}-\epsilon_{{\bf k}i}-\hbar\omega),
\end{equation}
\begin{equation}
\sigma_{xy}^{2} (\omega) = \frac{\pi e^2}{\hbar\omega m^2}
\sum_{i,j}\int_{BZ}\frac{d{\bf k}}{(2\pi)^3}\text{Im}[p_{ij}^{x}p_{ji}^{y}]
\delta(\epsilon_{{\bf k}j}-\epsilon_{{\bf k}i}-\hbar\omega),
\end{equation}
where \(\hbar\omega\) is the photon energy, 
and \(\epsilon_{\textbf{k}i(j)}\) are the energy eigenvalues of occupied (unoccupied) states. 
The transition matrix elements 
$p_{ij}^{a} = \langle\textbf{k}\emph{j}|\hat{p}_{a}|\textbf{k}i\rangle$ 
where \(|\textbf{k}i(j)\rangle\) 
are the $i$($j$)th occupied(unoccupied) states at 
$k$-point \(\textbf{k}\), and \(\hat{p}_a\) is the 
Cartesian component $a$ of the momentum operator. 
The imaginary part of the diagonal elements and the 
real part of the off-diagonal elements are then obtained 
from $\sigma_{aa}^{1} (\omega)$ and $\sigma_{xy}^{2} (\omega)$, 
respectively, via the Kramers-Kronig transformations as follows:
\begin{equation}
\sigma_{aa}^{2} (\omega) = -\frac{2\omega}{\pi }P \int _{0}^{\infty }\frac{\sigma_{aa}^{1}(\omega ')}{\omega^{'2}-\omega ^{2}}d\omega ^{'},
\end{equation}
\begin{equation}
\sigma_{xy}^{1} (\omega) = \frac{2}{\pi }P \int _{0}^{\infty }\frac{\omega^{'}\sigma_{xy}^{2}(\omega ')}{\omega^{'2}-\omega ^{2}}d\omega ^{'},
\end{equation}
where $P$ denotes the principal value of the integration. 
We can see that Eq. (2) and Eq. (3) neglect transitions across different $k$-points 
since the momentum of the optical photon is negligibly 
small compared with the electron crystal momentum
and thus only the direct interband transitions need to be considered. 
In our calculations \(p^a_{ij}\) are obtained in the PAW formalism~\cite{adolph2001}. 
We use a \(10 \times 10 \times  10\)  $k$-point mesh and the 
Brillouin zone integration is carried out 
with the linear tetrahedron method (see \cite{temmerman1989} and references therein), 
which leads to well converged results. 
To ensure that the $\sigma_{aa}^{2} (\omega)$ and $\sigma_{xy}^{1} (\omega)$ 
in the optical frequency range (e.g., $\hbar\omega < 8$ eV)
obtained via Eqs. (4) and (5) are converged,
we include the unoccupied states at least 21 eV above the Fermi energy, 
i.e., a total of 1200 (1300) bands are used in the YIG (BIG) calculations.

For a bulk magnetic material, the complex polar Kerr rotation angle is given by~\cite{guo1994,guo1995},
\begin{equation}
\theta _{K}+i\epsilon _{K}=\frac{-\sigma _{xy}}{\sigma _{xx}\sqrt{1+i(4\pi/\omega)\sigma _{xx}}}.
\end{equation}
Similarly, the complex Faraday rotation angle for a thin film can be written as~\cite{ravindran1999}
\begin{equation}
\theta _{F}+i\epsilon _{F}=\frac{\omega d}{2c}(n_{+}-n_{-}),
\end{equation}
where $n_+$ and $n_-$ represent the refractive indices for left- and right-handed polarized lights, respectively, 
and are related to the corresponding dielectric function (or optical conductivity via expressions 
$n_{\pm }^{2}=\varepsilon_{\pm}=1+{\frac{4\pi i}{\omega}}\sigma _{\pm}= 
1+{\frac{4\pi i}{\omega}}(\sigma _{xx}\pm i \sigma _{xy})$.
Here the real parts of the optical conductivity $\sigma _{\pm}$ can be written as
\begin{equation}
\sigma_{\pm}^{1} (\omega) = \frac{\pi e^2}{\hbar\omega m^2}
\sum_{i,j}\int_{BZ}\frac{d{\bf k}}{(2\pi)^3}|\Pi_{ij}^{\pm}|^{2}
\delta(\epsilon_{{\bf k}j}-\epsilon_{{\bf k}i}-\hbar\omega),
\end{equation}
where $\Pi_{ij}^{\pm} = \langle\textbf{k}\emph{j}|\frac{1}{\sqrt{2}}(\hat{p}_{x}\pm i\hat{p}_{y})|\textbf{k}i\rangle$. 
Clearly, $\sigma _{xy} = \frac{1}{2i}(\sigma _{+}-\sigma _{-})$, and this shows that $\sigma _{xy}$ 
would be nonzero only if $\sigma _{+}$ and $\sigma _{-}$ are different. 
In other words, magnetic circular dichroism is the fundamental cause of the
nonzero $\sigma _{xy}$ and hence the MO effects.  

\section{RESULTS AND DISCUSSION }
\subsection{Magnetic moments}

\begin{table*}[htbp]
\begin{center}
\caption{Total spin magnetic moment ($m_s^t$), atomic 
spin magnetic moments ($m_s^{Fe}$, $m_s^{O}$, $m_s^{Y(Bi)}$),
atomic Fe orbital magnetic moments ($m_o^{Fe}$) and band gap ($E_g$) 
of ferrimagnetic Y$_3$Fe$_5$O$_{12}$ and Bi$_3$Fe$_5$O$_{12}$ 
from the full-relativistic electronic structure calculations. 
For comparison, the available measured optical $E_g$ and 
total magnetization $m^t_{exp}$ are also listed.}
\begin{ruledtabular}
\begin{tabular}{ccccccc}
structure  & $m^t$ ($m^t_{exp}$) & $m_s^{Fe(16a)}$ ($m_o^{Fe(16a)}$) & $m_s^{Fe(24d)}$ ($m_o^{Fe(24d)}$) & $m_s^{O}$  &$m_s^{Y(Bi)}$ & $E_g$ ($E_g^{exp}$)    \\
  &  ($\mu_B$/f.u.) & ($\mu_B$/atom) & ($\mu_B$/atom) & ($\mu_B$/atom)   & ($\mu_B$/atom)& (eV) \\
 \hline
 Y$_3$Fe$_5$O$_{12}$ & 4.999 (5.0\footnotemark[1]) & -4.177 (-0.016) & 4.075 (0.018) & 0.067& 0.005 & 1.81 (2.4\footnotemark[2]) \\
 \hline
 Bi$_3$Fe$_5$O$_{12}$  & 4.996 (4.4\footnotemark[3])& -4.161 (-0.018) & 4.068 (0.019) &0.066 & 0.005 & 1.82 (2.1\footnotemark[4]) \\
\end{tabular}
\end{ruledtabular}
\footnotemark[1]{Ref.~\onlinecite{rodic1999}.}
\footnotemark[2]{Ref.~\onlinecite{wittekoek1975}.}
\footnotemark[3]{Ref.~\onlinecite{adachi2002}.}
\footnotemark[4]{Ref.~\onlinecite{jesenska2016}.}
\end{center}
\end{table*}
Here we first present calculated total and atom-decomposed magnetic moments in Table I.
As expected, Y$_3$Fe$_5$O$_{12}$ is a ferrimagnet 
in which Fe ions of the same type couple ferromagnetically
while Fe ions of different types couple antiferromagnetically. 
Since there are two  Fe$^O$ ions and three  Fe$^T$ ions in a unit cell, 
Y$_3$Fe$_5$O$_{12}$ is ferrimagnetic 
with a total magnetic moment per f.u. being $\sim$5.0 $ \mu_B$ (see Table I).
The calculated spin magnetic moments of Fe ions of both types are $\sim$4.0 $ \mu_B$,
being consistent with the high spin state of Fe$^{+2}$ 
($d^{5\uparrow}t_{2g}^{1\downarrow}$) ions in
either octahedral or tetrahedral crystal field. 
We note that the orbital magnetic moments of Fe 
are parallel to their spin magnetic moments.
Nonetheless, the calculated orbital magnetic moments of Fe are small, 
because of strong crystal field quenching.
Interestingly, there is a significant spin magnetic moment on each O ion, 
and this together with the spin magnetic moment of one net Fe 
ion per f. u. leads to the total spin magnetic moment
per f.u. of $\sim$5.0 $ \mu_B$.
The calculated Fe magnetic moments for  both symmetry sites 
agree rather well with the measured ones of $\sim4.0$ $ \mu_B$.~\cite{rodic1999}
The calculated total magnetization of $\sim5.0$ $\mu_B$/f.u. is 
also in excellent agreement with the experiment.~\cite{rodic1999} 

Bi$_3$Fe$_5$O$_{12}$ is also predicted to be ferrimagnetic, 
although the calculated magnetic moments 
of both Fe$^O$ and Fe$^T$ ions are slightly smaller than the corresponding 
ones in Y$_3$Fe$_5$O$_{12}$ (see Table I).
The total magnetization and local magnetic moments of the other ions in 
BIG are almost identical to that in YIG.
However, the experimental $m_{tot}$ for BIG is only $4.4$ $\mu_B$,~\cite{adachi2002} 
being significantly smaller than the calculated value. 
As mentioned before, stable high quality BIG crystals are
hard to grow. Consequently, this notable discrepancy in 
total magnetization between the calculation 
and the previous experiment~\cite{adachi2002} could be due to the 
poor quality of the samples used in the experiment.

\subsection{Electronic structure}
\begin{figure}
\centerline{\psfig{file=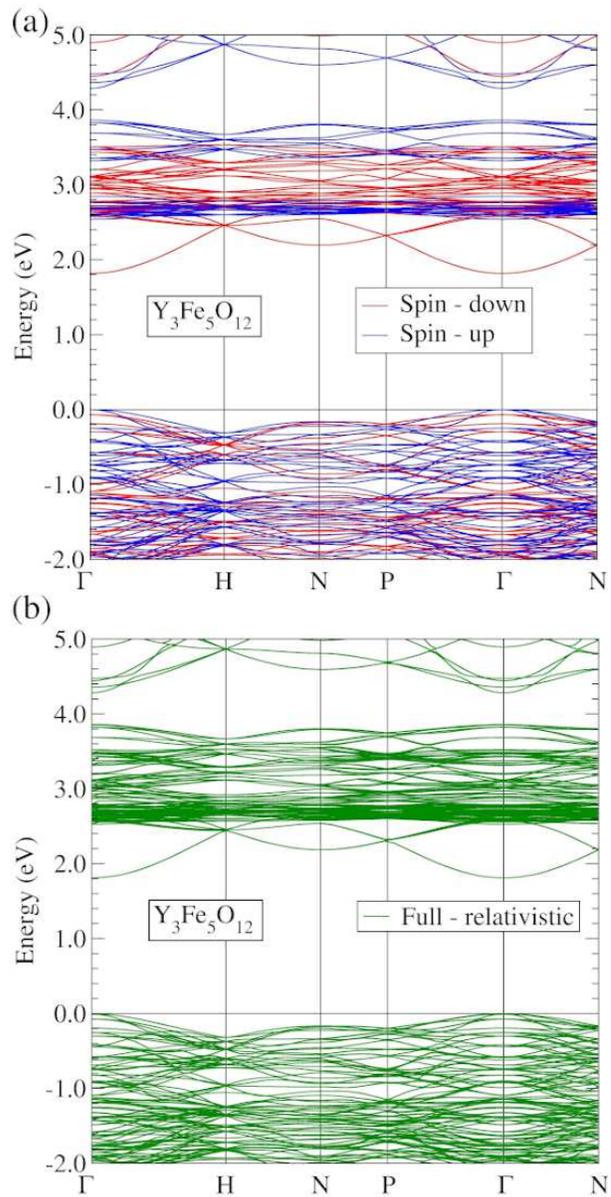,width=0.95\linewidth}}
\caption{(a) Scalar-relativistic spin-polarized band structure and (b) fully relativistic band structure of Y$_3$Fe$_5$O$_{12}$.}
\label{fig.2}
\end{figure}

 \begin{figure}
\centerline{\psfig{file=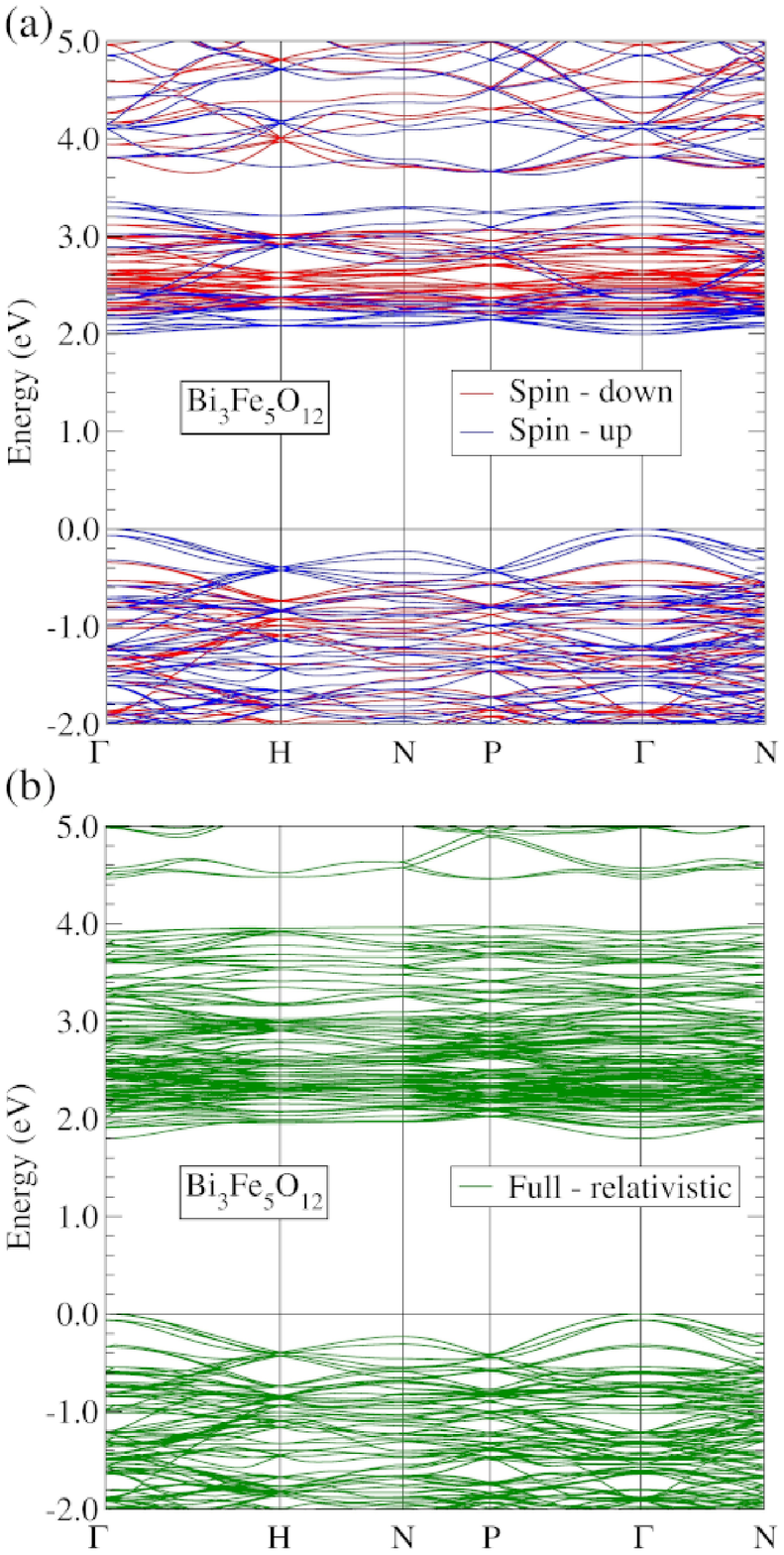,width=0.95\linewidth}}
\caption{(a) Scalar-relativistic spin-polarized band structure and (b) fully relativistic band structure of Bi$_3$Fe$_5$O$_{12}$.}
\label{fig.3}
\end{figure}

\begin{figure}
\centerline{\psfig{file=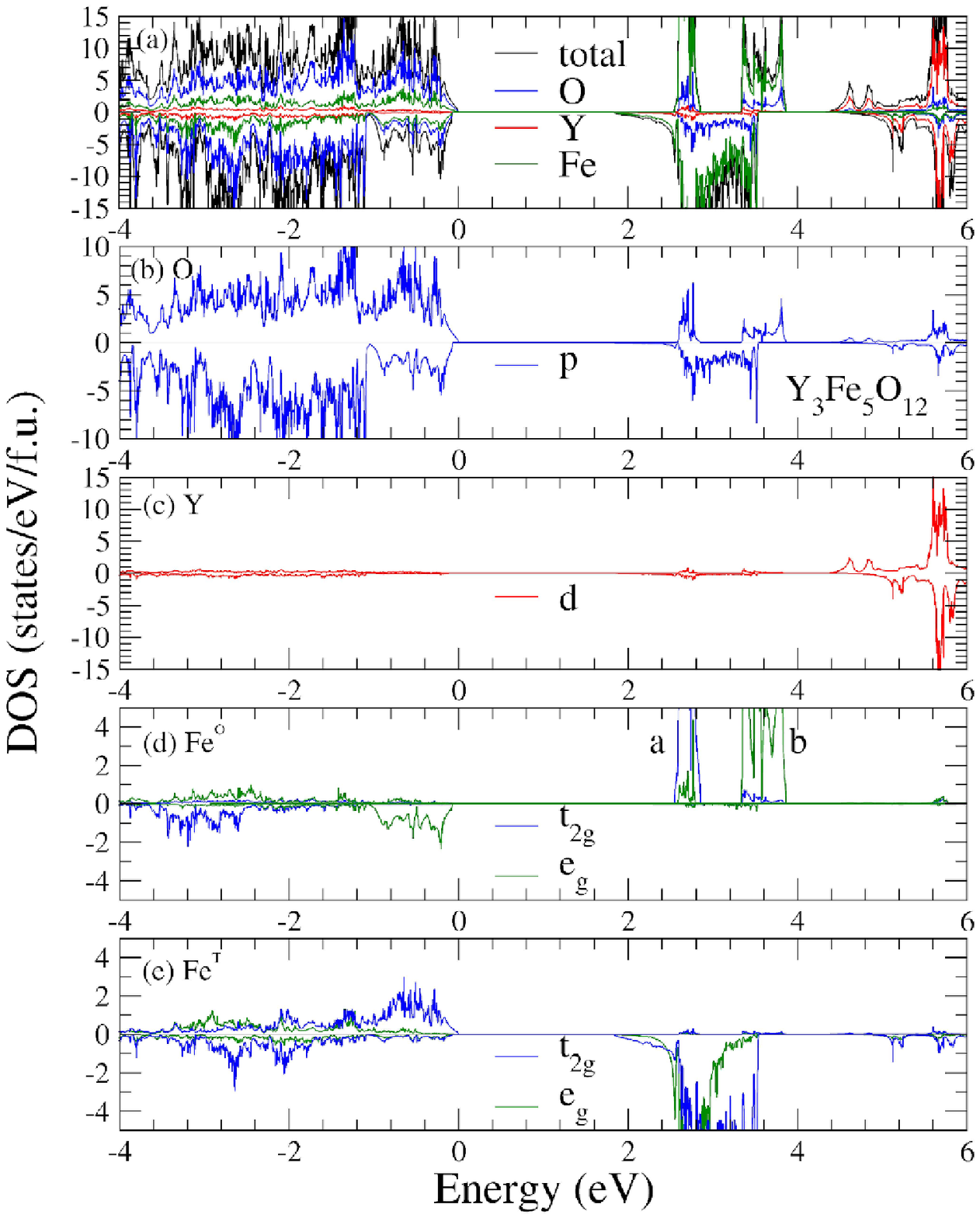,width=1.0\linewidth}}
\caption{Spin-polarized density of states (DOS) of Y$_3$Fe$_5$O$_{12}$ from the scalar-relativistic calculation.}
\label{fig.4}
\end{figure}

\begin{figure}
\centerline{\psfig{file=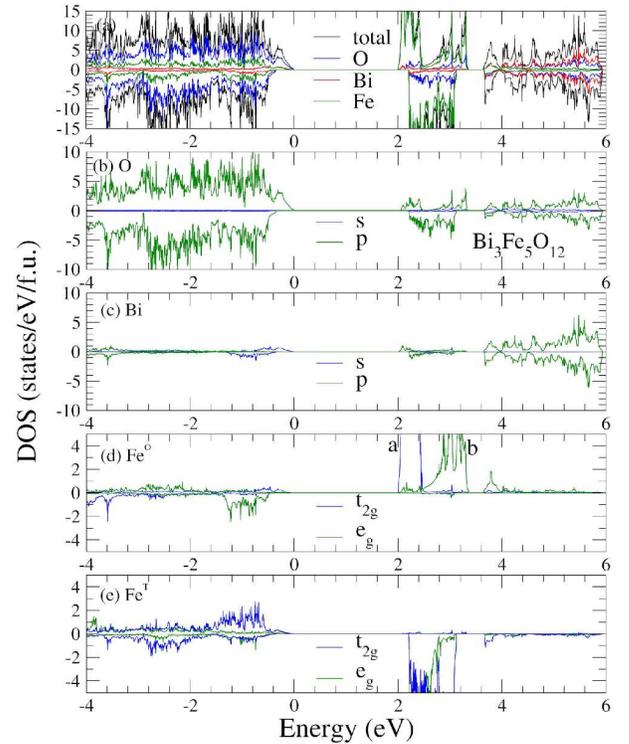,width=1.0\linewidth}}
\caption{Spin-polarized density of states (DOS) of Bi$_3$Fe$_5$O$_{12}$ from the scalar-relativistic calculation.}
\label{fig5}
\end{figure}

Here we present the calculated scalar-relativistic band structures of YIG and BIG 
in Fig. 2(a) and Fig. 3(a), respectively. 
The calculated band structures  show that YIG and BIG are 
both direct band-gap semiconductors, 
where the conduction band minimum (CBM) and valence band maximum (VBM) 
are both located at the $\Gamma$ point. 
For BIG, both CBM and VBM are purely spin-up bands. 
This means that BIG is a single-spin semiconductor, 
which may find applications for spintronic and spin photovoltaic devices. 
The origin of the MO effects is the magnetic circular dichorism [see Eq. (8)], as mentioned above,
which cannot occur without the presence of the spin-orbit coupling (SOC).
Therefore, it is useful to examine how the SOC influence the band structures.
The fully relativistic band structures for YIG and BIG are presented in Fig. 2(b) and Fig. 3(b), respectively.
First, we notice that with the inclusion of the SOC,
YIG and BIG are still direct band-gap semiconductors,
where the CBM and VBM are both located at the $\Gamma$ point.
Second, Fig. 3(b) indicates that when the SOC is considered, the BIG band structure changes
significantly, while the YIG band structure hardly changes [see Fig. 2(b)]. 
For example, the band gap for BIG decreases from 2.0 to 1.8 eV after the SOC is included.
Also, the gap, which was at 3.4 to 3.7 eV above the Fermi energy [see Fig. 3(a)],
now becomes from 3.9 to 4.5 eV above the Fermi energy [see Fig. 3(b)].
Interestingly, the substitution of yittrium by bismuth 
not only enhances the SOC but also changes the electronic band structure
significantly, as can be seen by comparing Figs. 2 and 3.

We also calculate total as well as site-, orbital-, and spin-projected 
densities of states (DOS) for YIG and BIG, as displayed in Fig. (4) and (5), respectively. 
First, Figs. (4) and (5) show that in both YIG and BIG, the upper valence bands ranging 
from -4.0 to 0.0 eV, are dominated by O $p$-orbitals with minor contributions from 
Fe $d$-orbitals as well as Y $d$-orbitals in YIG and Bi $sp$-orbitals in BIG.
Second, the lower conduction band manifold, ranging from 1.8 to $\sim$3.9 eV in YIG (Fig. 4)
and from 2.0 to 3.4 eV in BIG (Fig. 5), stems predominately from Fe $d$-orbitals with
small contributions from O $p$-orbitals.
Therefore, the semiconducting band gaps in YIG and BIG are mainly of the charge transfer type.
Furthermore, on the Fe$^T$ sites, the $d$-DOS in this conduction band is almost fully spin-down
[see Figs. 4(e) and 5(e)]. On the Fe$^O$ sites, on the other hand, the $d$-DOS
in this conduction band is almost purely spin-up [see Figs. 4(d) and 5(d)].
Here, the DOS peak marked $a$ mostly consists of $t_{2g}$ orbital 
while that marked $b$ above peak $a$, is made up of mainly $e_g$ orbital. 
The gap between peaks $a$ and $b$ is thus caused by the crystal field splitting. 

Figure 4 indicates that in YIG, the upper conduction bands from 4.4 to 6.0 eV are mainly of Y $d$ orbital character
with some contribution from O $p$ orbitals. In BIG, on the other hand, the upper conduction bands from 3.6 to 6.0 eV 
are mainly the Bi and O $p$ orbital hybridized bands (see Fig. 5). 
Notably, there is sizable Bi $sp$ DOS in the lower conduction band region from 2.0 to 3.4 eV (see Fig. 5(c)],
indicating that the lower conduction bands in BIG are significantly mixed with Bi $sp$ orbitals,
as noticed already by Oikawa {\it et al.}~\cite{oikawa2005},
Since the SOC of the Bi $p$ orbitals are very strong, this explains why the band width of 
the lower conduction bands in BIG increases from $\sim$1.4 to 2.1 eV when the SOC is included (see Fig. 3).
In contrast, the band width of the lower conduction bands in YIG remains unaffected by the SOC (see Fig. 2).
This also explains why the MO effects in BIG are much stronger than in YIG, as reported in Sec. III.D. below.

\begin{figure}
\centerline{\psfig{file=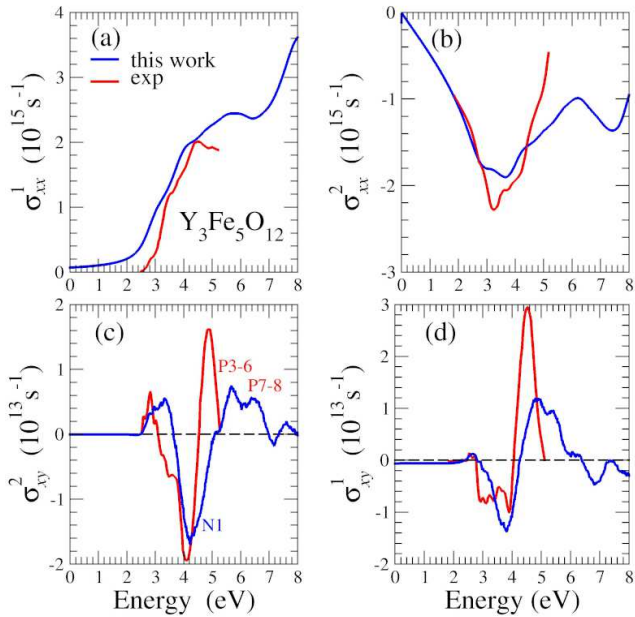,width=1\linewidth}}
\caption{Calculated optical conductivity of Y$_3$Fe$_5$O$_{12}$. (a) Real part and (b) imaginary part 
of the diagonal element; (c) imaginary part and (d) real part of the off-diagonal element.
All the spectra have been convoluted with a Lorentzian of 0.3 eV to simulate the finite quasiparticle lifetime effects. 
Red lines are the optical conductivity derived from the experimental dielectric constant.~\cite{wittekoek1975}}
\label{fig.6}
\end{figure}

\begin{figure}
\centerline{\psfig{file=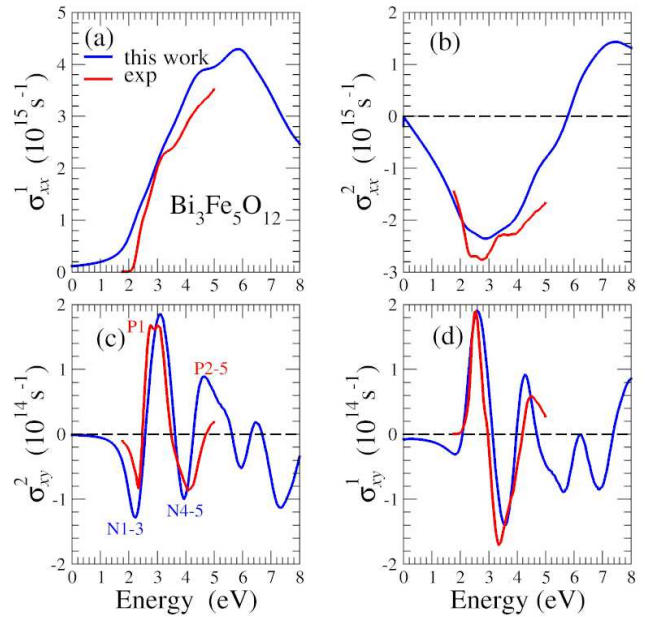,width=1\linewidth}}
\caption{Calculated optical conductivity of Bi$_3$Fe$_5$O$_{12}$. (a) Real part and (b) imaginary part
of the diagonal element; (c) imaginary part and (d) real part of the off-diagonal element.
All the spectra have been convoluted with a Lorentzian of 0.3 eV to simulate the finite quasiparticle lifetime effects.
Red lines are the optical conductivity derived from the experimental dielectric constant.~\cite{jesenska2016}
}
\label{fig.7}
\end{figure}
 
\subsection{Optical Conductivity}

Here we present the optical and magneto-optical conductivities for YIG and BIG which
are ingredients for calculating the Kerr and Faraday rotation angles [see Eq. (6) and Eq. (7)].
In particular, the MO conductivity (i.e., the off-diagonal element of the conductivity tensor $\sigma_{xy}$)
is crucial, as shown by Eq. (8). 
Calculated optical conductivity spectra of YIG and BIG are plotted as a function of photon energy
in Fig. 6 and Fig. 7, respectively. 
For YIG, the real part of the diagonal element of the conductivity tensor ($\sigma^1_{xx}$) starts to increase rapidly 
from the absorption edge ($\sim$ 2.3 eV) to $\sim$ 4.0 eV, and then further increases with a smaller slope 
up to $\sim$ 5.6 eV [see Fig. 6(a)]. It then decreases slightly until 6.6 eV and finally increases again with a much steeper 
slope up to $\sim$ 8.0 eV. Similarly, in BIG, \(\sigma^1_{xx}\) increases steeply from the absorption edge ($\sim$ 2.0 eV)  
to  $\sim$ 4.0 eV, and then further increases with a smaller slope up to $\sim$ 6.0 eV [see Fig. 7(a)]. 
It then decrease steadily from $\sim$ 6.0 eV to $\sim$ 8.0 eV. 
The behaviors of the imaginary part of the diagonal element (\(\sigma^2_{xx}\))
of YIG and BIG are rather similar in the energy range up to 5.0 eV [see Figs. 6(b) and 7(b)]. 
The \(\sigma^2_{xx}\) spectrum has a broad valley at $\sim$ 3.5 eV ($\sim$ 3.0 eV) in the case of YIG (BIG). 
However, the \(\sigma^2_{xx}\) spectra of YIG and BIG differ from each other for energy $>$ 5.0 eV. 
There is a sign change in \(\sigma^2_{xx}\) occuring at $\sim$ 5.8 eV for BIG, 
while there is no such a sign change in \(\sigma^2_{xx}\) of YIG up to 8.0 eV.

The striking difference in the  off-diagonal element of the conductivity ($\sigma_{xy}$) 
(i.e., magneto-optical conductivity or magnetic circular dichroism)
between YIG and BIG is that $\sigma_{xy}$ of BIG is almost ten times larger than that of YIG (see Figs. 6 and 7). 
Nonetheless, the line shapes of the off-diagonal element of YIG and BIG
are rather similar except that their signs seem to be opposite and their peaks appear at quite different energy positions.
In particular, in the low energy range up to $\sim$4.4 eV,  
the line shape of the imaginary part of the off-diagonal element (\(\sigma^2_{xy}\)) 
of BIG looks like a "W" [see Fig. 7(c)], while that of YIG in the energy region up to $\sim$7.0 eV
seems to have the inverted "W" shape [see Fig. 6(c)], The main difference
is that the \(\sigma^2_{xy}\) of BIG decreases oscillatorily from 4.4 to 8.0 eV.
On the other hand, the line shape of the real part of the off-diagonal element (\(\sigma^1_{xy}\)) 
of BIG looks like a "sine wave" between 2.0 and 4.7 eV [see Fig. 7(d)],
while that of YIG appears to be an inverted "sine wave" between 2.6 and 6.4 eV [see Fig. 6(d)].
The largest magnitude of \(\sigma^2_{xy}\) of YIG is $\sim 1.6 \times10^{13}s^{-1}$ at $\sim$ 4.3 eV,
while that of BIG is $\sim 1.9 \times10^{14}s^{-1}$ at $\sim$ 3.1 eV.
The largest magnitude of \(\sigma^1_{xy}\) of YIG is $\sim 1.2 \times10^{13}s^{-1}$ at $\sim$ 4.8 eV,
while that of BIG is $\sim 1.9 \times10^{14}s^{-1}$ at $\sim$ 2.6 eV.

 
In order to compare with the available experimental data, 
we also plot the experimental optical conductivity spectra~\cite{wittekoek1975,jesenska2016} in Figs. 6 and 7.
The theoretical spectra of the diagonal element of the optical conductivity tensor for both 
YIG and BIG match well with that of the experimental ones in the measured energy range 
[see Figs. 6(a) and 6(b) as well as Figs. 7(a) and 7(b)]. 
Interestingly, we note that the relativistic GGA+U calculations give rise to 
the band gaps of YIG and BIG that are smaller than the experimental ones (see Table II),
and yet the calculated and measured optical spectra agree rather well with each other.
This apparently contradiction can be resolved as follows.
In YIG, for example, the lowest conduction bands at $E = 1.8 \sim 2.4$ eV above the VBM
are highly dispersive (see Fig. 2) and thus have very low DOS (see Fig. 4).
This results in very low optical transition.
Therefore, the main absoption edge that appears in the optical spectrum ($\sigma_{xx}^1$)
is $\sim2.2$ eV, which is close to the experimental absorption edge of 2.5 eV,
instead of 1.8 eV as determined by the calculated band structure (see Table II).
In contrast, no such highly dispersive bands appear at the CBM in BIG,
Thus the calculated band gap agrees better with the measured band gap~\cite{jesenska2016} (Table II).

Figures 7(c) and 7(d) show that the calculated $\sigma_{xy}^1$ and $\sigma_{xy}^2$ of BIG
agree almost perfectly with the experimental data~\cite{jesenska2016}. 
The peak positions, peak heights and overall trend of the theoretical spectra
are nearly identical to that of the experimental ones~\cite{jesenska2016}.
On the other hand, the calculated $\sigma_{xy}^1$ and $\sigma_{xy}^2$ 
for YIG do not agree so well with the experimental data~\cite{wittekoek1975} [Figs. 6(c) and 6(d].
For example, there is a sharp peak at $\sim$ 4.8 eV 
in the experimental $\sigma_{xy}^1$ spectrum, which seems to be shifted to
a higher energy at 5.6 eV with much reduced magnitude in the theoretical $\sigma_{xy}^1$ spectrum [see Fig. 6(c)]. 
Also, for $\sigma_{xy}^2$ spectrum, there is a sharp peak at $\sim$ 4.5 eV in the experimental $\sigma_{xy}^2$ spectrum,
which appears at $\sim4.8$ eV with considerably reduced height [see Fig. 6(d)].
Nonetheless, the overall trend of the theoretical $\sigma_{xy}$ spectra
of YIG is in rather good agreement with that of the measured ones~\cite{wittekoek1975}. 

Equations (2), (3), and (8) indicate that the absorptive parts of the optical conductivity
elements ($\sigma_{xx}^1, \sigma_{zz}^1, \sigma_{xy}^2$ and $\sigma_{\pm}^1$) are 
directly related to the dipole allowed interband transitions.
Thus, we analyze the origin of the main features in the magneto-optical conductivity ($\sigma_{xy}^2$)
spectrum by determining the symmetries of the involved band states and the dipole selection rules
(see the Appendix for details). The absorptive optical spectra are usually dominated
by the interband transitions at the high symmetry points where the energy bands are generally 
flat (see, e.g., Figs. 2 and 3), thus resulting in large joint density of states.
As an example, here we consider the interband optical transitions 
at the $\Gamma$ point where the band extrema often occur.
Based on the determined band state symmetries and dipole selection rules 
(see Table III in the Appendix) as well as calculated transition matrix elements [Im$(p_{ij}^xp_{ji}^y)$],
we assign the main features in $\sigma_{xy}^2$ [labelled in Figs. 6(c) and 7(c)] 
to the main interband transitions at the $\Gamma$ point as shown in Figs. 8 and 9.
The details of these assignments, related interband transitions and transition matrix elements
for YIG and BIG are presented in Tables IV and V in the Appendix, respectively.
Since there are too many possible transitions to list, we present only those transitions
whose transition matrix elements $|$Im$(p_{ij}^xp_{ji}^y) | > 0.010 $ a.u. in YIG (Table IV) and
$|$Im$(p_{ij}^xp_{ji}^y) | > 0.012$ a.u. in BIG (Table V).

Figure 8 shows that nearly all the main optical transitions in YIG 
are from the upper valence bands to the upper conduction bands,
and only one main transition (P$_3$) to the lower conduction bands. 
Consequently, these transitions contribute to the main features in $\sigma_{xy}^2$ 
at photon energy $> 4.0$ eV [see Fig. 6(c)].
In contrast, in BIG, a large number of the main transitions (e.g., P1-5, P7, N1-4, N5-8) 
are from the upper valence bands to lower conduction bands (see Fig. 9).
This gives rise to the main features in $\sigma_{xy}^2$ for photon energy $< 4.0$ eV
[see Fig. 7(c)], whose magnitudes are generally one order of magnitude larger than
that of  $\sigma_{xy}^2$ in YIG, as mentioned above. 
The largely enhanced MO activity in BIG stems from the significant hybridization
of Bi $p$-orbitals with Fe $d$-orbitals in the lower conduction bands, as mentioned above.
Since heavy Bi has a strong spin-orbit coupling, this hybridization greatly increases
the dichroic interband transitions from the upper valence bands to the lower 
conduction bands in BIG. As mentioned above, Y $sd$ orbitals contribute significantly
only to the upper conduction bands in YIG, and this results in the pronounced
magneto-optical transitions only from the upper valence bands to the upper conduction
bands (Fig. 8). Furthermore, Y is lighter than Bi and thus has a weaker SOC than Bi. 

The discussion in the proceeding paragraph clearly indicates that the significant
hybridization of heavy Bi $p$ orbitals with Fe $d$ orbitals
in the lower conduction bands just above the band gap
is the main reason for the large MO effect in BIG.
The magnetism in BIG is mainly caused by the iron $d$ orbitals 
which have a rather weak SOC. However, through the hybridization
between Bi $p$ orbitals and Fe $d$ orbitals, the strong SOC
effect is also transfered to the lower conduction bands.
Large exchange splitting and strong
spin-orbit coupling in the valence and conduction bands
below and above the band gap are crucial
for strong magnetic circular dichroism and hence large MO effects.
Therefore, in search of materials with strong MO effects,
one should look for magnetic systems that contain heavy elements such as Bi
and Pt~\cite{guo1996}. 
\begin{figure}
\centerline{\psfig{file=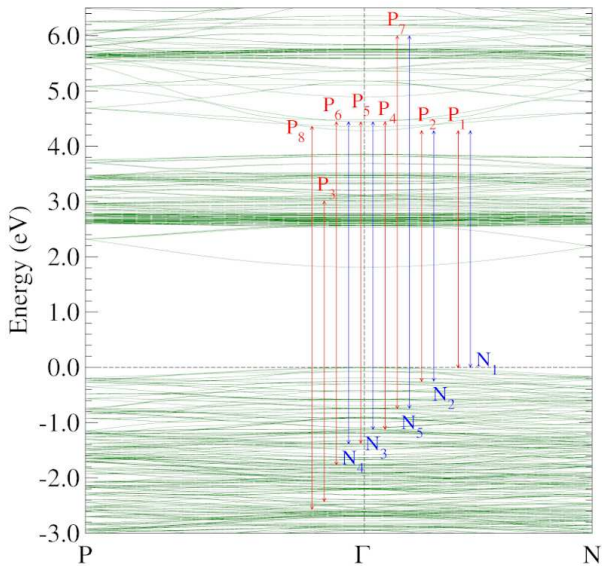,width=0.95\linewidth}}
\caption{Relativistic band structures of Y$_3$Fe$_5$O$_{12}$. Horizontal dashed lines denote the top of valance band. 
The principal interband transitions at the $\Gamma$ point and the corresponding peaks in the $\sigma_{xy}$ in Fig. 6 (c) 
are indicated by red and blue arrows.}
\label{fig.8}
\end{figure}

\begin{figure}
\centerline{\psfig{file=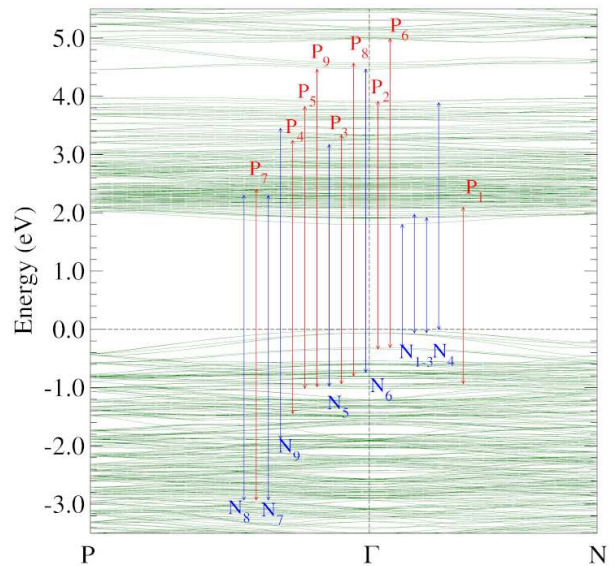,width=0.95\linewidth}}
\caption{Relativistic band structures of Bi$_3$Fe$_5$O$_{12}$. Horizontal dashed lines denote the top of valance band. 
The principal interband transitions at the $\Gamma$ point and the corresponding peaks in the $\sigma_{xy}$ in Fig. 7 (c) 
are indicated by red and blue arrows.}
\label{fig.9}
\end{figure}



\subsection{Magneto-optical Kerr and Faraday effect }
Finally, let us study the polar Kerr and Faraday effects in YIG and BIG. 
The complex Kerr and Faraday rotation angles for YIG and BIG are plotted 
as a function of photon energy in Figs. 8 and 9, respectively. 
First of all, we notice that the Kerr rotation angles of BIG [Fig. 10(c)]
are many times larger than that of YIG [Fig. 10(a)].
For example, the positive Kerr rotation maximum of 0.10 \degree$ $ in YIG
occurs at $\sim3.6$ eV, while that (0.80 \degree) for BIG appears at $\sim3.5$ eV. 
The negative Kerr rotation maximum (-0.12 \degree) of YIG occurs at $\sim4.8$ eV,
while that (-1.21 \degree) for BIG appears at $\sim2.4$ eV.
This may be expected because Kerr rotation angle is proportional to the
MO conductivity ($\sigma_{xy}^{1}$) [Eq. (6)], which in BIG is nearly
ten times larger than in YIG, as mentioned in the proceeding subsection. 
Similarly, the Kerr ellipticity maximum (0.16 \degree) of YIG occurs at $\sim4.1$ eV [Fig. 10(b)],
whereas that (0.54 \degree) of BIG [Fig. 10(d)] appears at $\sim1.9$ eV. 
The negative Kerr ellipticity maximum (-0.07 \degree) of YIG occurs at $\sim5.7$ eV
while that (-1.16\degree) of BIG is located at $\sim2.9$ eV.

Let us now compare our calculated Kerr rotation angles with some known MO materials 
such as $3d$ transition metal alloys and compound semiconductors.~\cite{antonov2004}
For magnetic metals, ferromagnetic $3d$ transition metals and their alloys are an important family. 
Among  them, manganese-based pnictides are known to have strong MO effects. 
In particular, MnBi thin films were reported to have a large Kerr rotation angle of 2.3 \degree.~\cite{di1996,ravindran1999}
Platinum alloys such as FePt, Co$_2$Pt ~\cite{guo1996} and PtMnSb~\cite{vanengen1983} also
possess large Kerr rotation angles. 
It was shown that the strong SOC on heavy Pt in these systems is the main cause of the strong MOKE. ~\cite{guo1996} 
Among semiconductor MO materials, diluted magnetic semiconductors 
Ga$_{1-x}$Mn$_x$As were reported to show Kerr rotations angle 
as large as 0.4 \degree at 1.80 eV.~\cite{lang2005} 
Therefore, the strong MOKE effect in YIG and BIG could 
have promising applications in high density MO data-storage devices 
or MO nanosensors with high spatial resolution. 

Figure 9 shows that as for the Kerr rotation angles,
the Faraday rotation angles of BIG are generally up to ten times larger than
that of YIG. The Faraday rotation maximum (7.2 \degree/$\mu$m) of YIG
occurs at $\sim3.9$ eV, while that (51.2\degree/$\mu$m) of BIG is located at $\sim3.7$ eV.
The Faraday ellipticity maximum (7.9 \degree/$\mu$m) for YIG appears at $\sim4.4$ eV,
whereas that (54.1\degree/$\mu$m) of BIG occurs at $\sim2.3$ eV.
On the other hand, the negative Faraday rotation maximum (-5.7\degree/$\mu$m)
occurs at $\sim5.4$ eV, while that (-74.6\degree/$\mu$m) for BIG appears at $\sim2.7$ eV. 
The negative Faraday ellipticity maximum (-3.6\degree/$\mu$m) of YIG occurs at $\sim6.6$ eV, 
while that (-70.2\degree/$\mu m$) for BIG is located at $\sim3.2$ eV. 
For comparision, we notice that MnBi films are known to possess large Faraday rotation angles 
of $\sim 80$\degree/$\mu$m at 1.8 eV.~\cite{di1996,ravindran1999} 

Finally, we compare our predicted MOKE and MOFE spectra with the available experiments in Figs. 10 and 11. 
All the predicted MOKE and MOFE spectra are in rather good agreement with the experimental 
ones in the experimental photon energy range~\cite{kahn1969,wittekoek1975,jesenska2016,deb2012}. 
Nonetheless, our theoretical predictions would have a better agreement with the experiments if all the 
calculated spectra are blue-shifted slightly by $\sim$0.3 eV, thus suggesting that the
theoretical band gaps are slightly too small. 

\begin{figure}
\centerline{\psfig{file=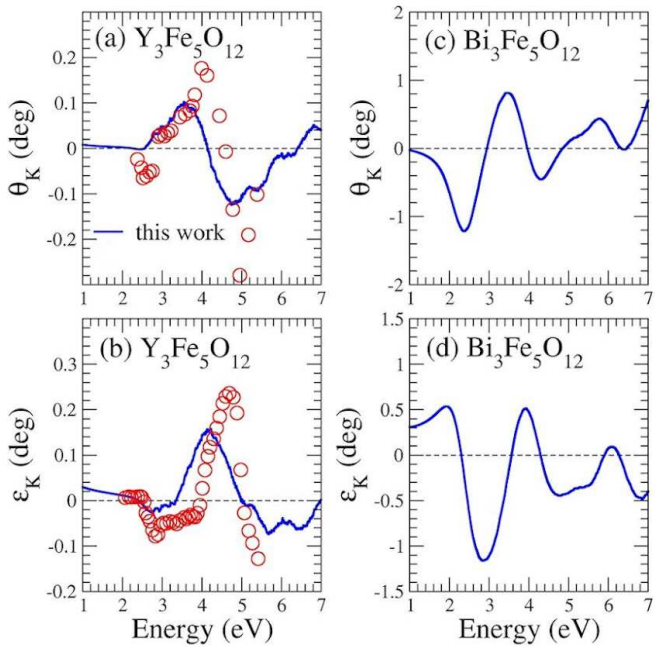,width=1.05\linewidth}}
\caption{Calculated complex Kerr rotation angles (blue curves). (a) Kerr rotation ($\theta_K$) 
and (b) Kerr ellipticity ($\varepsilon_K$) spectra 
of Y$_3$Fe$_5$O$_{12}$; (c) Kerr rotation ($\theta_K$) and (d) Kerr ellipticity ($\varepsilon_K$) spectra
of Bi$_3$Fe$_5$O$_{12}$.  Red circles in (a) and (b) denote the experimental values from Ref. ~\cite{kahn1969}.}
\label{fig.10}
\end{figure}

\begin{figure}
\centerline{\psfig{file=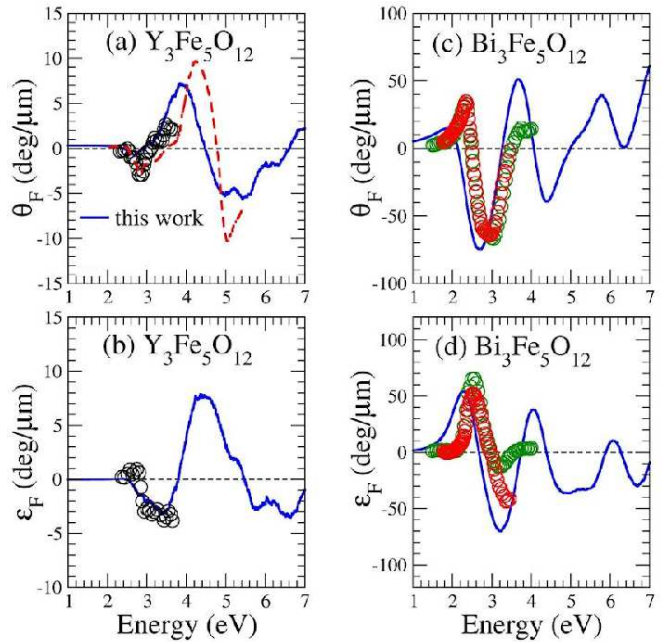,width=1.05\linewidth}}
\caption{Calculated complex Faraday rotation angles (blue curves). 
(a) Faraday rotation ($\theta_F$) and 
(b) Faraday ellipticity ($\varepsilon_F$) spectra of Y$_3$Fe$_5$O$_{12}$; 
(c) Kerr rotation ($\theta_F$) and 
(d) Kerr ellipticity ($\varepsilon_F$) spectra
of Bi$_3$Fe$_5$O$_{12}$. 
Red dashed line in (a) denotes the measured values from Ref. ~\cite{kahn1969}. 
Black circles in (a) and (b) are the experimental values from Ref. ~\cite{wittekoek1975}. 
Red (green) circles in (c) and (d) are the experimental values 
from Ref. ~\cite{deb2012} (\cite{jesenska2016}) } 
\label{fig.11}
\end{figure}

\section{Conclusion}
To summarize, we have systematically studied the 
electronic structure, magnetic, optical and MO properties 
of cubic iron garnets YIG and BIG
by performing GGA+U calculations. 
We find that YIG exhibits significant MO Kerr and Faraday effects
in UV frequency range  that are comparable to cubic ferromagnetic iron.
Strikingly,  we find that BIG shows gigantic MO effects in
the visible frequency region that are several times larger than YIG. 
In particular, the Kerr rotation angle of BIG 
becomes as large as -1.2\degree at photon energy 2.4 eV, 
and the Faraday rotation angle for the BIG film reaches 
-75 \degree/$\mu m$ at 2.7 eV.  
Calculated MO conductivity ($\sigma_{xy}^2$) spectra reveal that
these distinctly different MO properties of YIG and BIG result
from the fact that the magnitude of $\sigma_{xy}^2$ of BIG
is nearly ten times larger than that of YIG.
Our calculated Kerr and Faraday rotation angles of YIG agree well with
the available experimental values. Our calculated Faraday rotation
angles of BIG are in nearly perfect agreement with the measured ones.
Thus, we hope that our predicted giant MO Kerr effect in BIG
will stimulate further MOKE experiments on high quality BIG crystals.` 


Principal features in the optical and MO spectra 
are analyzed in terms of the calculated band structures
especially the symmetry of the band states and optical transition matrix elements 
at the $\Gamma$ point of the BZ. 
We find that in YIG, Y $sd$ orbitals mix mainly with the upper conduction
bands that are $\sim4.5$ eV above the VBM, and thus leave
the Fe $d$ orbital dominated lower conduction bands from 1.8 to 3.8 eV
above the VBM almost unaffected by the SOC on the Y atom.
In contrast, Bi $p$ orbitals in BIG hybridize significantly
with Fe $d$ orbitals in the lower conduction bands and this 
leads to large SOC-induced band splitting and much increased band width
of the lower conduction bands. Consequently, the MO transitions 
between the upper valence bands and lower conduction bands 
are greatly enhanced when Y is replaced by heavier Bi.
This finding thus provides a guideline in search for materials with
desired MO effects, i.e., one should look for magnetic materials
with heavy elements such as Bi whose orbitals hybridize
significantly with the MO active conduction or valence bands.

Finally, our findings of strong MO effects in these iron garnets 
and also single-spin semiconductivity in BIG 
suggest that cubic iron garnets are an useful playground of exploring the 
interplay of microwave, spin current, magnetism, and optics degrees of freedom,
and also have promising applications in high density semiconductor MO data-storage
and low-power consumption spintronic nanodevices.

\section*{Acknowledgments}
The authors thank Ming-Chun Jiang for many valuable discussions throughout this work.
The authors acknowledge the support from the Ministry of Science and Technology and the National
Center for Theoretical Sciences (NCTS) of The R.O.C. The authors are also grateful to the National
Center for High-performance Computing (NCHC) for the computing time.
G.-Y. Guo also thanks the support from the Far Eastern Y. Z. Hsu Science and Technology
Memorial Foundation in Taiwan.

\section*{APPENDIX: Dipole selection rules and symmetries of band states at $\Gamma$}
In this Appendix, to help identify the origins of the main features in the magneto-optical conductivity
$\sigma_{xy}(\omega)$ spectra of YIG and BIG, we provide the dipole selection rules
and the symmetries of the band states at the $\Gamma$ as well as the main optical transitions between them.

Both YIG and BIG have the \textit{Ia$\bar{3}$d} space group
and thus they have the $C_{4h}$ ($4/mm'm'$) point group at the $\Gamma$ point in the Brillouin zone. 
Based on the character table of the $C_{4h}$ point group~\cite{koster1963},
we determine the dipole selection rules for the optical transitions between the band states at the $\Gamma$ point, as listed in Table III. 
We calculate the eigenvalues for all symmetry elements of each eigenstate of the $\Gamma$ point using 
the \textit{Irvsp} program~\cite{gao2020} and then determine the irreducible representation and hence the symmetry of the state.
Based on the obtained symmetries of the band states and also calculated optical matrix elements [Im$(p_{ij}^xp_{ji}^y)$]
[see Eq. (3)], we assign the peaks in the $\sigma_{xy}(\omega)$ spectra of YIG [see Fig. 6(c)] and BIG [see Fig. 7(c)]
to the main optical transitions at the $\Gamma$ point (see Fig. 8 and 9, respectively), as listed in Tables IV and V, respectively.

\begin{table}
\caption{Dipole selection rules for the $C_{4h}$ point group at the $\Gamma$ point in the Brillouin zone of YIG and BIG.
}
\begin{tabular}{llllllllll}
 &  &  &  &  &  &  &  &  &  \\\hline\hline
\multicolumn{1}{l|}{polarization} & $\Gamma_6^+$ & $\Gamma_5^+$ & $\Gamma_8^+$ & $\Gamma_7^+$ & $\Gamma_6^-$ & $\Gamma_5^-$ & $\Gamma_8^-$ & $\Gamma_7^-$ &  \\\hline
\multicolumn{1}{l|}{$z$} & $\Gamma_6^-$ & $\Gamma_5^-$ & $\Gamma_8^-$ & $\Gamma_7^-$ & $\Gamma_6^+$ & $\Gamma_5^+$ & $\Gamma_8^+$ & $\Gamma_7^+$ &  \\
\multicolumn{1}{l|}{$x+iy$} & $\Gamma_5^-$ & $\Gamma_8^-$ & $\Gamma_7^-$ & $\Gamma_6^-$ & $\Gamma_5^+$ & $\Gamma_8^+$ & $\Gamma_7^+$ & $\Gamma_6^+$ &  \\
\multicolumn{1}{l|}{$x-iy$} & $\Gamma_7^-$ & $\Gamma_6^-$ & $\Gamma_5^-$ & $\Gamma_8^-$ & $\Gamma_7^+$ & $\Gamma_6^+$ & $\Gamma_5^+$ & $\Gamma_8^+$ &  \\\hline\hline
\end{tabular}

\end{table}

\begin{table}
\caption{Main optical transitions between the states at the $\Gamma$ point of the Brillouin zone of YIG. 
Symbols in the first column denote the assigned peaks in the magneto-optical conductivity ($\sigma_{xy}^{2}$) spectrum
(Figs. 6 and 8). $i$ and $j$ denote the initial and final states, respectively.
Im$(p_{ij}^xp_{ji}^y)$ denote the calculated transition matrix element (in atomic units) [see Eq. (3)].
$E_i$ and $E_j$ represent the initial and final state energies (in eV),
respectively. $\Delta E_{ij} = E_j-E_i$ is the transition energy.}
\begin{tabular}{ccccccc}
\hline \hline
Peak   & state $i$    &  state $j$  & Im$(p_{ij}^xp_{ji}^y)$ & $\Delta E_{ij}$ & $E_j$  & $E_i$  \\ \hline
P8 & 545 ($\Gamma_6^+$) &  802 ($\Gamma_5^-$) &  0.0103 &  6.928 &  4.358 &  -2.569 \\
P3 & 556 ($\Gamma_5^-$) &  761 ($\Gamma_8^+$) &  0.0115 &  5.438 &  3.010 &  -2.428 \\
P6 & 609 ($\Gamma_7^-$) &  803 ($\Gamma_6^+$) &  0.0102 &  6.207 &  4.442 &  -1.765 \\
N4 & 632 ($\Gamma_5^-$) &  803 ($\Gamma_6^+$) &  -0.0273 &  5.825 &  4.442 &  -1.384 \\
P5 & 635 ($\Gamma_7^-$) &  803 ($\Gamma_6^+$) &  0.0251 &  5.818 &  4.442 &  -1.376 \\
N3 & 649 ($\Gamma_5^-$) &  803 ($\Gamma_6^+$) &  -0.0146 &  5.567 &  4.442 &  -1.125 \\
P4 & 651 ($\Gamma_7^-$) &  803 ($\Gamma_6^+$) &  0.0195 &  5.564 &  4.442 &  -1.123 \\
P7 & 668 ($\Gamma_6^+$) &  844 ($\Gamma_5^-$) &  0.0136 &  6.744 &  5.994 &  -0.750 \\
N5 & 670 ($\Gamma_8^+$) &  844 ($\Gamma_5^-$) &  -0.0119 &  6.739 &  5.994 &  -0.745 \\
P2 & 690 ($\Gamma_6^-$) &  801 ($\Gamma_5^+$) &  0.0116 &  4.537 &  4.280 &  -0.257 \\
N2 & 692 ($\Gamma_8^-$) &  801 ($\Gamma_5^+$) &  -0.0107 &  4.535 &  4.280 &  -0.254 \\
P1 & 698 ($\Gamma_6^-$) &  801 ($\Gamma_5^+$) &  0.0431 &  4.292 &  4.280 &  -0.012 \\
N1 & 700 ($\Gamma_8^-$) &  801 ($\Gamma_5^+$) &  -0.0452 &  4.280 &  4.280 &  0.000 \\\hline\hline
\end{tabular}
\end{table}

\begin{table}
\caption{Main optical transitions between the states at the $\Gamma$ point of the Brillouin zone of BIG.
Symbols in the first column denote the assigned peaks in the magneto-optical conductivity ($\sigma_{xy}^{2}$) spectrum
(Figs. 7 and 9). $i$ and $j$ denote the initial and final states, respectively. 
Im$(p_{ij}^xp_{ji}^y)$ denote the calculated transition matrix element (in atomic units) [see Eq. (3)].
$E_i$ and $E_j$ represent the initial and final state energies (in eV),
respectively. $\Delta E_{ij} = E_j-E_i$ is the transition energy.}
\begin{tabular}{ccccccc}
\hline\hline
Peak   &   state $i$    &   state $j$    & Im$(p_{ij}^xp_{ji}^y)$ & $\Delta E_{ij}$ & $E_j$  & $E_i$  \\ \hline
N8 & 578 ($\Gamma_8^+$) & 783 ($\Gamma_5^-$) & -0.0130 & 5.234 & 2.304 & -2.930 \\
P7 & 578 ($\Gamma_8^+$) & 789 ($\Gamma_7^-$) &   0.0123 & 5.331 & 2.401 & -2.930 \\
N7 & 579 ($\Gamma_6^+$) & 782 ($\Gamma_7^-$) & -0.0139 & 5.228 & 2.298 & -2.930 \\
N9 & 661 ($\Gamma_6^-$) & 856 ($\Gamma_7^+$) & -0.0136 & 5.333 & 3.452 & -1.882 \\
P4 & 684 ($\Gamma_7^-$) & 846 ($\Gamma_6^+$) &  0.0127 & 4.685 & 3.242 & -1.443 \\
P5 & 709 ($\Gamma_8^+$) & 869 ($\Gamma_7^-$) &  0.0139 & 4.839 & 3.825 & -1.014 \\
P9 & 710 ($\Gamma_6^+$) & 873 ($\Gamma_5^-$) &  0.0149 & 5.453 & 4.467 & -0.986 \\
N5 & 712 ($\Gamma_8^+$) & 842 ($\Gamma_5^-$) & -0.0121 & 4.153 & 3.175 & -0.979 \\
P3 & 715 ($\Gamma_7^+$) & 854 ($\Gamma_6^-$) &  0.0135 & 4.271 & 3.336 & -0.935 \\
P8 & 723 ($\Gamma_7^+$) & 876 ($\Gamma_6^-$) &  0.0142 & 5.374 & 4.571 & -0.803 \\
N6 & 727 ($\Gamma_8^+$) & 873 ($\Gamma_5^-$) & -0.0145 & 5.206 & 4.467 & -0.738 \\
P2 & 741 ($\Gamma_6^-$) & 872 ($\Gamma_5^+$) &  0.0122 & 4.254 & 3.917 & -0.337 \\
P6 & 743 ($\Gamma_5^-$) & 878 ($\Gamma_8^+$) &  0.0127 & 5.305 & 4.991 & -0.314 \\
N3 & 743 ($\Gamma_5^-$) & 749 ($\Gamma_6^+$) & -0.0184 & 2.113 & 1.799 & -0.314 \\
N2 & 744 ($\Gamma_8^-$) & 755 ($\Gamma_5^+$) & -0.0160 & 2.037 & 1.973 & -0.063 \\
N1 & 745 ($\Gamma_7^-$) & 753 ($\Gamma_8^+$) & -0.0142 & 1.980 & 1.917 & -0.062 \\
N4 & 747 ($\Gamma_7^-$) & 871 ($\Gamma_8^+$) & -0.0138 & 3.891 & 3.891 & 0.000 \\
P1 & 715 ($\Gamma_7^+$) & 760 ($\Gamma_6^-$) &  0.0125 & 3.032 & 2.097 & -0.935 \\\hline\hline
\end{tabular}
\end{table}


\begin{thebibliography}{10}
\bibitem{cherepanov1993}V. Cherepanov, I. Kolokolov, and V. L’vov, 
The saga of YIG: spectra, thermodynamics, interaction and 
relaxation of magnons in a complex magnet, Phys. Rep. {\bf 229}, 81 (1993).

\bibitem{mizukami2002}S. Mizukami, Y. Ando, and T. Miyazaki, 
Effect of spin diffusion on Gilbert damping for a very thin permalloy layer 
in Cu/permalloy/Cu/Pt films, Phys. Rev. B{\bf 66}, 104413 (2002).

\bibitem{chikazumi1997}S. Chikazumi, \textit{Physics of Ferromagnetism}, 
2nd ed. (Oxford University Press, Oxford, 1997)

\bibitem{kajiwara2010} Y. Kajiwara, K. Harii, S. Takahashi, J. Ohe, K. Uchida, 
M. Mizuguchi, H. Umezawa, H. Kawai, K. Ando, K. Takanashi, S. Maekawa, and E. Saitoh, 
Transmission of electrical signals by spin-wave interconversion in a 
magnetic insulator, Nature (London) {\bf 464}, 262 (2010).

\bibitem{schneider2008}T. Schneider, A. A. Serga, B. Leven, B. Hillebrands, 
R. L. Stamps, and M. P. Kostylev, Realization of spin-wave logic gates, 
Appl. Phys. Lett. {\bf 92}(2), 022505 (2008).

\bibitem{sun2013}Y. Sun, H. Chang, M. Kabatek, Y.-Y. Song, Z. Wang, 
M. Jantz, W. Schneider, M. Wu, E. Montoya, B. Kardasz, B. Heinrich,
S. G. E. te Velthuis, H. Schultheiss, and A. Hoffmann, 
Damping in Yttrium Iron Garnet Nanoscale Films Capped by Platinum, 
Phys. Rev. Lett. {\bf 111}, 106601 (2013).

\bibitem{oppeneer2001} P. M. Oppeneer, 
Chapter 1 Magneto-optical Kerr Spectra, pp. 229-422, 
in 
\newblock {\em Handbook of Magnetic Materials}, edited by K. H. J. Buschow.
\newblock Elsevier, Amsterdam, (2001).








\bibitem{antonov2004} V.~Antonov, B.~Harmon, and A.~Yaresko.
\newblock {\em Electronic structure and magneto-optical properties of solids}.
\newblock Springer Science \& Business Media,  (2004).


\bibitem{castera1996} J. P. Castera, in {\it Magneto-optical Devices}, Vol. {\bf 9} of 
{\it Encyclopedia of Applied Physics}, edited by G. L. Trigg (Wiley-VCH, New York, 1996), p. 133.

\bibitem{mansuripur1995} M. Mansuripur, 
{\it The Principles of Magneto-Optical Recording} (Cambridge Univ. Press, Cambridge, 1995).



\bibitem{haldane2008}F. D. M. Haldane and S. Raghu, 
Possible Realization of Directional Optical Waveguides in Photonic Crystals 
with Broken Time-Reversal Symmetry, Phys. Rev. Lett. {\bf 100}, 013904 (2008).

\bibitem{aplet1964}L. J. Aplet and J. W. Carson, A Faraday effect optical isolator, 
Appl. Opt. {\bf 3}, 544 (1964).

\bibitem{dillon1958}J. F. Dillon,  Optical properties of several ferrimagnetic garnets, 
J. Appl. Phys. {\bf 29}, 539 (1958).


\bibitem{wittekoek1975}S. Wittekoek, T. J. A. Popma, J. M. Robertson, P. F. Bongers, 
Magneto-optic spectra and the dielectric tensor elements of 
bismuth-substituted iron garnets at photon energies between 2.2-5.2 eV, 
Phys. Rev. B {\bf12}, 2777 (1975).

\bibitem{kahn1969}F. J. Kahn, P. S. Pershan, and J. P. Remeika, 
Ultraviolet Magneto-Optical Properties of Single-Crystal Orthoferrites, 
Garnets, and Other Ferric Oxide Compounds, Phys. Rev. {\bf186}, 891 (1969).

\bibitem{chern1999} M. -Y. Chern, F. -Y. Lo, D. -R. Liu, K. Yang, and J. -S. Liaw, 
Red shift of Faraday rotation in thin films of completely 
bismuth-substituted iron garnet Bi$_3$Fe$_5$O$_{12}$
, Jpn. J. App. Phys., Part 1 {\bf 38}, 6687 (1999).

\bibitem{jesenska2016}E. Jesenska, T. Yoshida, K. Shinozaki, T. Ishibashi, L. Beran, 
M. Zahradnik, R. Antos, M. Ku$\check{\textup{c}}$era, and M. Veis, Optical and magneto-optical properties 
of Bi substituted yttrium iron garnets prepared by metal organic decomposition, 
Opt. Mater. Express {\bf6}(6), 1986-1997 (2016).


\bibitem{vertruyen2008} B. Vertruyen, R. Cloots, J. S. Abell, T. J. Jackson, 
R. C. da Silva, E. Popova, and N. Keller, Curie temperature, 
exchange integrals, and magneto-optical properties in off-stoichiometric 
bismuth iron garnet epitaxial films, Phys. Rev. B {\bf 78}, 094429 (2008).

\bibitem{xu2000}Y. -N. Xu, Z. -Q. Gu, and W. Y. Ching, 
First-principles calculation of the electronic structure of yttrium iron garnet 
(Y$_3$Fe$_5$O$_{12}$), J. Appl. Phys. {\bf 87}, 4867 (2000).
%
	
\bibitem{oikawa2005}T. Oikawa, S. Suzuki, and K. Nakao, 
First-principles study of spin-orbit interactions in 
bismuth iron garnet, J. Phys. Soc. Jpn. {\bf 74}, 401 (2005).
%

\bibitem{bertaut1956}F. Bertaut, F. Forrat, A. Herpin, 
and P. M\'eriel, \' Etude par diffraction de neutrons du grenat 
ferrimagn\'etique Y$_3$Fe$_5$O$_{12}$, Compt. rend. {\bf 243}, 898 (1956).

\bibitem{toraya1995}H. Toraya and T. Okuda, Crystal structure analysis 
of polycrystalline Bi$_3$Fe$_5$O$_{12}$ thin film by using asymmetric 
and symmetric diffraction techniques, J. Phys. Chem. Solids {\bf56}, 1317 (1995).

\bibitem{perdew1996}
J.~P. Perdew, K.~Burke, and M.~Ernzerhof, Generalized
Gradient Approximation Made Simple, Phys. Rev. Lett.{ \bf 77}, 3865 (1996).

\bibitem{dudarev1998} S. L. Dudarev, G. A. Botton, S. Y. Savrasov, 
C. J. Humphreys and A. P. Sutton, Electron-energy-loss spectra 
and the structural stability of nickel oxide: An LSDA+ U study, 
Phys. Rev. B{ \bf 57}, 1505 (1998).

\bibitem{jeng2004} H.-T. Jeng, G. Y. Guo and D. J. Huang, 
Charge-orbital ordering and Verwey transition in magnetite, 
Phys. Rev. Lett.{ \bf 93}, 156403 (2004).

\bibitem{kresse1999}
G.~Kresse and D.~Joubert, From ultrasoft pseudopotentials 
to the projector augmented-wave method, Phys. Rev. B {\bf 59}, 1758 (1999).

\bibitem{kresse1996a}
G.~Kresse and J.~Furthm{\"u}ller, Efficient iterative schemes
for \textit{ab initio} total-energy calculations using a plane-wave
basis set, Phys. Rev. B {\bf 54}, 11169 (1996).

\bibitem{kresse1996b}
G.~Kresse and J.~Furthm{\"u}ller, Efficiency of 
ab-initio total energy calculations for metals and 
semiconductors using a plane-wave basis set, 
Comput. Mat. Sci {\bf 6}, 15 (1996).


\bibitem{feng2015} W. Feng, G.-Y. Guo, J. Zhou, 
Y. Yao, and Q. Niu, Large magneto-optical Kerr effect 
in noncollinear antiferromagnets 
Mn$_3X$ ($X=$ Rh, Ir, Pt) Phys. Rev. B {\bf 92}, 144426 (2015).

\bibitem{wang1974} C. S. Wang and J. Callaway, 
Band structure of nickel: Spin-orbit coupling, the Fermi surface, 
and the optical conductivity, Phys. Rev. B {\bf 9}, 4897 (1974).

\bibitem{oppeneer1992} P. M. Oppeneer, T. Maurer, J. Sticht, 
and J. Kübler, $Ab$ $initio$ calculated magneto-optical Kerr effect 
of ferromagnetic metals: Fe and Ni, Phys. Rev. B {\bf 45}, 10924 (1992).


\bibitem{adolph2001} B. Adolph, J. Furthmüller, and F. Bechstedt, 
Optical properties of semiconductors using projector-augmented waves, 
Phys. Rev. B {\bf 63}, 125108 (2001).

\bibitem{temmerman1989} W. M. Temmerman, P. A. Sterne, 
G. Y. Guo, and Z. Szotek, Electronic Structure Calculations of High T$_c$
Materials, Mol. Simul. {\bf 63}, 153 (1989).

\bibitem{guo1994} G.-Y.~Guo and H.~Ebert, Theoretical investigation 
of the orientation dependence of the magneto-optical Kerr effect in 
Co, Phys. Rev. B {\bf 50}, 10377 (1994).

\bibitem{guo1995} G.-Y.~Guo and H.~Ebert, Band-theoretical investigation 
of the magneto-optical Kerr effect in Fe and Co multilayers, 
Phys. Rev. B {\bf 51}, 12633 (1995).

\bibitem{rodic1999}D. Rodic, M. Mitric, R. Tellgren, H. Rundlof, and 
A. Kremenovic, True magnetic structure of the ferrimagnetic garnet 
Y$_3$Fe$_5$O$_{12}$ and magnetic moments of iron ions,
 J. Magn. Magn. Mater. {\bf191}, 137 (1999).

\bibitem{adachi2002}N. Adachi, T. Okuda, V. P. Denysenkov, 
A. Jalali-Roudsar, and A. M. Grishin, Magnetic properties of single crystal film 
Bi$_3$Fe$_5$O$_{12}$ prepared onto Sm$_3$(Sc,Ga)$_5$O$_{12}$(1 1 1), 
J. Magn. Magn. Mater. 242-245, 775 (2002).





\bibitem{guo1996} G. Y.~Guo and H.~Ebert, On the origins
of the enhanced magneto-optical Kerr effect in ultrathin Fe and Co
multilayers, J. Magn. Magn. Mater. \textbf{156}, 173 (1996).

\bibitem{ravindran1999} P.~Ravindran, A.~Delin, P.~James,
B.~Johansson, J.~Wills, R.~Ahuja, and O.~Eriksson, Magnetic, optical,
and magneto-optical properties of MnX (X=As, Sb, or Bi) from full-potential calculations,
Phys. Rev. B{ \bf 59}, 15680 (1999).

\bibitem{di1996} G. Q.~Di and S.~Uchiyama, 
Optical and magneto-optical properties of MnBi film, 
Phys. Rev. B{ \bf 53}, 3327 (1996).


\bibitem{vanengen1983} P.~Van~Engen, K.~Buschow, R.~Jongebreur, 
and M.~Erman, PtMnSb, a material with very high magneto-optical Kerr effect, Appl. Phys. Lett. {\bf 42}, 202--204 (1983).

\bibitem{lang2005}  R. Lang, A. Winter, H. Pascher, H. Krenn, X. Liu, 
and J. K. Furdyna, Polar Kerr effect studies of Ga$_{1-x}$Mn$_x$As 
epitaxial films, Phys. Rev. B{ \bf 72}, 024430 (2005).


\bibitem{deb2012}M. Deb, E. Popova, A. Fouchet, and N. Keller, 
Magneto-optical Faraday spectroscopy of completely 
bismuth-substituted Bi$_3$Fe$_5$O$_{12}$ garnet thin films, J. Phys. D {\bf45}, 455001 (2012).

\bibitem{koster1963}G. F. Koster, J. O. Dimmock, R. G. Wheeler, and H. Statz, Properties of the thirty-two point groups (Vol. 24). MIT press (1963).

\bibitem{gao2020}J.-C. Gao, Q.-S. Wu, C. Persson, and Z.-J. Wang, Irvsp: to obtain irreducible representations of electronic states in the VASP, arXiv preprint arXiv:2002.04032 (2020).








\end{thebibliography}
\end{document}